\documentclass[journal]{IEEEtran}
\iffalse
  \usepackage[mode=buildnew]{standalone}
\else
  \newcommand\includestandalone[1]{\includegraphics{#1}}
\fi
\usepackage{graphicx}
\usepackage[lighttt]{lmodern}
\usepackage{listings}
\usepackage[plain]{algorithm}
\usepackage{algpseudocode}
\usepackage{verbatim}
\usepackage{pgfplots}
\usepackage{tabularx}
\usepackage{float}
\usepackage[utf8]{inputenc}
\usepackage[T1]{fontenc}
\usetikzlibrary{plotmarks}

\newlength{\figurewidth} 
\lstdefinelanguage [GCC]{C++} []{C++} {
  morekeywords={[2]llong},%
  morekeywords={[3]__syncthreads,__device__,__global__}%
}
\newcommand{\ie}{{\it i.e. }}
\newcommand{\code}[1]{\lstinline!#1!}

\begin{document}

\title{Benchmarking the cost of thread divergence in CUDA}
\author{P.~Bialas and A.~Strzelecki%
\thanks{Authors are with Faculty of Physics, Astronomy and Computer Science, %
 Jagiellonian University, ul. Łojasiewicza 11, 30-348 Krakow, Poland}
}
\date{\today}

\maketitle


\begin{abstract}

All modern processors include a set of vector instructions. While this gives a
tremendous boost to the performance, it requires a vectorized code that can
take advantage of such instructions. As an ideal vectorization is hard to
achieve in practice, one has to decide when different instructions may be
applied to different elements of the vector operand. This is especially
important in implicit vectorization as in NVIDIA CUDA Single Instruction
Multiple Threads (SIMT) model, where the vectorization details are hidden from
the programmer. In order to assess the costs incurred by incompletely
vectorized code, we have developed a micro-benchmark that measures the
characteristics of the CUDA \emph{thread divergence} model on different
architectures focusing on the loops performance.

\end{abstract}

\begin{IEEEkeywords}
GPU, CUDA, multithreading, SIMT, SIMD
\end{IEEEkeywords}


\section{Introduction}

Most of the current processors derive their performance from some form of SIMD
instructions. This holds true for \emph{Intel} i-Series and \emph{Xeon}
processor (8 lanes wide AVX instruction), \emph{Intel Xeon Phi} accelerator (16
lanes wide AVX instructions), \emph{AMD Graphics Cores Next} (64 lanes wide
wavefronts) and \emph{NVIDIA CUDA} (32 lanes wide warps). By the very nature of
those instructions all the operations on a vector are performed in parallel, at
least conceptually. That provides a constraint on the class of algorithms that
can be efficiently implemented on such architectures. If we want to perform
different operations on the different components of the vector operands we are
essentially forced to issue different instructions masking the unused
components in each of them which of course carries a performance penalty. The
aim of this paper is to asses those cost. We will concentrate on the NVIDIA
CUDA as that is the architecture that we use in our current
work\cite{Reconstruction}.

The CUDA and OpenCL programming models make \emph{thread divergence} very easy
to achieve as they rely on \emph{implicit} vectorization. They model the
computing environment as a great number of threads executing a single program
called \emph{kernel}\/\cite{Tesla}, and to the programmer this may look as an
multithreaded parallel execution. In fact the CUDA Programming Guide states:
``For the purposes of correctness, the programmer can essentially ignore the
SIMT behavior; however, substantial performance improvements can be realized by
taking care that the code seldom requires threads in a warp to diverge.''

On NVIDIA architectures threads are grouped warps of 32 threads that must all
execute same instructions in essentially SIMD (vector) fashion. This entails
that any branching instruction with condition that does not give the same
result across the whole warp leads to thread divergence - while some threads
take one branch other do nothing and then the other branch is taken with roles
reversed. This is a picture given in the NVIDIA programming guide, which warns
against thread divergence but otherwise is missing any details. Only in the
Tesla architecture white-paper it is mentioned that this is achieved using a
\emph{branch synchronization stack} \cite{Tesla}.

As the strict avoidance of thread divergence would severely limit the
algorithms that could be implemented using CUDA, it would be interesting to
check what are the real costs. The necessity of following two branches is one
obvious performance obstacle. We will be however interested in the overhead
associated with the \emph{re-convergence} mechanism itself as manifested in
loops.

Loops, which are probably the most used control statement in programming, are an
interesting example. While it is easy to picture the two way branching model as
in a \texttt{if else} statement, it is much harder to follow the execution of
loops, especially nested. In this contribution we will benchmark the
performance of single and double loops with bounds that are different across
the threads. While such model is discouraged in CUDA programming it can
nevertheless appear naturally while porting an existing multi-core algorithm to
GPU. To our best knowledge such measurements were not published up to this
date. We have only found some data in reference \cite{Demystifying} but not the
explicit timings of the synchronization stack operations. We will analyze our
results in view of the best information on the CUDA thread divergence model
that we have found.

This paper is organized as follows: In the next section we present the setup we
used for benchmarking the execution of single and double loop kernel. Then we
present the CUDA stack based re-convergence mechanism and in the last
section we use it to analyze our results.


\section{Timings}
\label{sec:timings}

\subsection{Single loop}

Our first test consisted of running single loop using the kernel from
Listing~\ref{lst:single-loop}. Each thread of a warp (we used only one warp)
runs through the same loop but with upper limit, denoted by $M$, set
individually for each thread. We measured the number of cycles taken by the
loop using the CUDA \lstinline!clock64()! function, repeating our loop
\lstinline!N_UNROLL! times if needed. We have also added a possibility to run
same loop \lstinline!N_PREHEAT! times before making the measurements. We have
repeated each measurement 256 times and used those samples to estimate the
error. We have found out that when using \lstinline!N_PREHEAT=1! we had
essentially zero errors even with \lstinline!N_UNROLL=1! . When no "preheating"
was used the results were more erratic, with difference of few cycles between
the samples.

We start our measurements with all loops having same upper limit of $M=32$ (no
divergence). Next we decrease the upper limit for one of the threads and
continue until all threads in warp have different upper bounds (see
Table~\ref{tab:single-loop}).

\begin{lstlisting}[linerange={7-8,11-39},label={lst:single-loop},float,
caption={Single loop kernel.}]
#define EXPR_INNER 1.3333f
#define EXPR_OUTER 2.3333f

__global__ void single_loop(int* limits, float* out, llong* timer) {

  int tid = blockDim.x * blockIdx.x + threadIdx.x;
  int M = limits[threadIdx.x];
  float sum = out[tid];

#pragma unroll
  for (int k = 0; k < N_PREHEAT; k++)
    for (int i = 0; i < M; i++) {
      sum += EXPR_INNER;
    }

  __syncthreads();
  llong start = clock64();

#pragma unroll
  for (int k = 0; k < N_UNROLL; ++k)
    for (int i = 0; i < M; i++) {
      sum += EXPR_INNER;
    }

  llong stop = clock64();
  __syncthreads();

  out[tid] = sum;
  timer[2 * tid] = start;
  timer[2 * tid + 1] = stop;
}
\end{lstlisting}

\begin{table*}[t]
\setlength{\tabcolsep}{3.55pt}
\begin{tabularx}{\textwidth}{|c|cccccccccccccccccccccccccccccccc|}
\hline
   & \multicolumn{32}{c|}{\tt tid}\\\hline
$n$ & 0 & 1 & 2 & 3 & 4 & 5 & 6 & 7 & 8 & 9 & 10 & 11 & 12 & 13 & 14 & 15 & 16 & 17 & 18 & 19 & 20 & 21 & 22 & 23 & 24 & 25 & 26 & 27 & 28 & 29 & 30 & 31 \\\hline
   & \multicolumn{32}{c|}{M}\\\hline
0  & 32 & 32 & 32 & 32 & 32 & 32 & 32 & 32 & 32 & 32 & 32 & 32 & 32 & 32 & 32 & 32 & 32 & 32 & 32 & 32 & 32 & 32 & 32 & 32 & 32 & 32 & 32 & 32 & 32 & 32 & 32 & 32\\
1  & 32 & 32 & 32 & 32 & 32 & 32 & 32 & 32 & 32 & 32 & 32 & 32 & 32 & 32 & 32 & 32 & 32 & 32 & 32 & 32 & 32 & 32 & 32 & 32 & 32 & 32 & 32 & 32 & 32 & 32 & 32 & 31\\
2  & 32 & 32 & 32 & 32 & 32 & 32 & 32 & 32 & 32 & 32 & 32 & 32 & 32 & 32 & 32 & 32 & 32 & 32 & 32 & 32 & 32 & 32 & 32 & 32 & 32 & 32 & 32 & 32 & 32 & 32 & 31 & 30\\
$\vdots$&\multicolumn{32}{|c|}{$\vdots$} \\
28 & 32 & 32 & 32 & 32 & 31 & 30 & 29 & 28 & 27 & 26 & 25 & 24 & 23 & 22 & 21 & 20 & 19 & 18 & 17 & 16 & 15 & 14 & 13 & 12 & 11 & 10 & 9 & 8 & 7 & 6 & 5 & 4\\
29 & 32 & 32 & 32 & 31 & 30 & 29 & 28 & 27 & 26 & 25 & 24 & 23 & 22 & 21 & 20 & 19 & 18 & 17 & 16 & 15 & 14 & 13 & 12 & 11 & 10 & 9 & 8 & 7 & 6 & 5 & 4 & 3\\
30 & 32 & 32 & 31 & 30 & 29 & 28 & 27 & 26 & 25 & 24 & 23 & 22 & 21 & 20 & 19 & 18 & 17 & 16 & 15 & 14 & 13 & 12 & 11 & 10 & 9 & 8 & 7 & 6 & 5 & 4 & 3 & 2\\
31 & 32 & 31 & 30 & 29 & 28 & 27 & 26 & 25 & 24 & 23 & 22 & 21 & 20 & 19 & 18 & 17 & 16 & 15 & 14 & 13 & 12 & 11 & 10 & 9 & 8 & 7 & 6 & 5 & 4 & 3 & 2 & 1\\
\hline
\end{tabularx}
\vspace{0.5em}
\caption{\label{tab:single-loop}Loop limits used in the single loop measurements.}
\end{table*}

\begin{figure}
\centering
\begin{tikzpicture}
\begin{axis}[
  width=\columnwidth,
  xlabel={$n$},
  ylabel={cycles},
  grid=major,
  xtick={0,4,8,12,16,20,24,28,32},
  legend style= {
    legend pos=north west,
    legend cell align=left
  }
]

\addplot+[ blue, only marks, mark=*, mark options={ fill=white } ]
table {data/ss_cuda_gtx470.txt};
\addlegendentry{Fermi}

\addplot+[ red, only marks, mark=square*, mark options={ fill=white } ]
table[x index = 0, y index = 5] {data/ss_cuda_gtx770_p1u001.txt};
\addlegendentry{Kepler}

\addplot+[ black, only marks, mark=triangle*, mark options={ fill=white } ]
table {data/ss_cuda_gtx850M.txt};
\addlegendentry{Maxwell}

\addplot+[red, smooth, no marks, domain=0:31] {1732+32*x};

\end{axis}
\end{tikzpicture}

\caption{\label{fig:single-loop}Single loop timings corresponding to kernel in
Listing~\ref{lst:single-loop}. }

\end{figure}
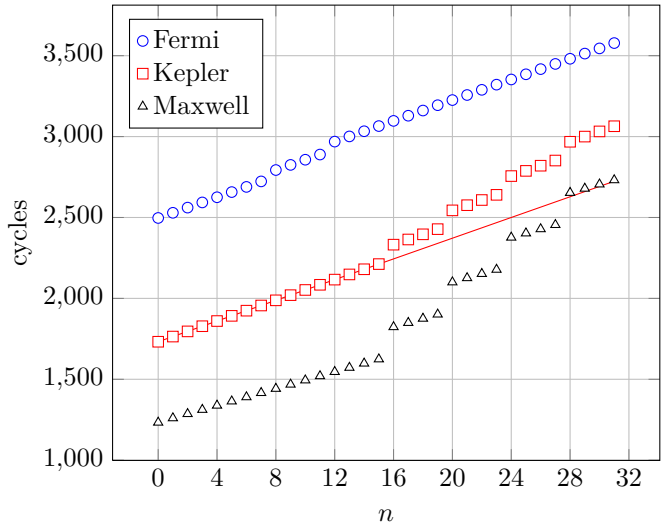

The results of the measurements as a function of $n$ are presented in the
Figure~\ref{fig:single-loop}. We have tested three CUDA architectures (see
Table~\ref{tab:arch}). We have found out that the number of cycles depended
only on the architecture or Compute Capability of the card, not on the
particular device. We have used CUDA 7.0 environment for all our test. We have
compiled our benchmarks using the \hbox{\lstinline!-arch=sm_20 -Xptxas -O3!
flags}, as \lstinline!sm_20! architecture did not change the results, but
produced no undefined instructions in the \lstinline!cuobjdump! disassembly
listing, contrary to \lstinline!sm_30!.

There are two interesting things on this plot to take notice of, apart from the
fact that GPU are getting faster with each new architecture. Firstly although
the longest loop has always same upper bound ($M=32$) the time increases
linearly with each new divergent thread. This is a clear signal of an
additional cost associated with the divergence, as in fact there is altogether
\emph{less} work done by the threads. Secondly this cost of thread divergence
is proportional to the number of divergent threads, up to $n=15$, then it
exhibits several "jumps" every four steps, at least for the two newest
architectures (\emph{Kepler}, \emph{Maxwell}). The behavior on the \emph{Fermi}
architecture is slightly different, but we will be not concerned with this
architecture in this contribution, and provide the plots only for comparison.

\begin{table}
\centering
\begin{tabular}{|l|l|}
\hline\hline
\multicolumn{1}{|c|}{architecture} & \multicolumn{1}{c|}{device} \\\hline
Fermi   & GTX 480,  GT 610\\
Kepler  & GT  650M, GT 755M, GTX 770\\
Maxwell & GTX 850M \\\hline
\hline
\end{tabular}

\caption{\label{tab:arch}Devices used in test.}

\end{table}

\subsection{Double loop}

\begin{lstlisting}[linerange={7-8,40-74}, label={lst:double-loop},  float,
caption={Double loop kernel.}]
#define EXPR_INNER 1.3333f
#define EXPR_OUTER 2.3333f

__global__ void double_loops(int* limits, float* out, llong* timer) {

  int tid = blockDim.x * blockIdx.x + threadIdx.x;
  int M = limits[2 * threadIdx.x];
  int N = limits[2 * threadIdx.x + 1];
  float sum = out[0];

#pragma unroll
  for (int k = 0; k < N_PREHEAT; k++)
    for (int i = 0; i < M; i++) {
      for (int j = 0; j < N; j++) {
        sum += EXPR_INNER;
      }
      sum += EXPR_OUTER;
    }

  __syncthreads();
  llong start = clock64();

#pragma unroll
  for (int k = 0; k < N_UNROLL; ++k)
    for (int i = 0; i < M; i++) {
      for (int j = 0; j < N; j++) {
        sum += EXPR_INNER;
      }
      sum += EXPR_OUTER;
    }

  llong stop = clock64();
  __syncthreads();
  out[tid] = sum;
  timer[2 * tid] = start;
  timer[2 * tid + 1] = stop;
}
\end{lstlisting}

\begin{figure}
\centering
\begin{tikzpicture}
\begin{axis}[
  width=\columnwidth,
  xlabel={$n$},
  ylabel={cycles},
  grid=major,
  xtick={0,4,8,12,16,20,24,28,32},
  legend style= {
    legend pos=north west,
    legend cell align=left
  }
]

\addplot+[ blue, only marks, mark=*, mark options={ fill=white } ]
table {data/ds_cuda_gtx470.txt};
\addlegendentry{Fermi}

\addplot+[ red, only marks, mark=square*, mark options={ fill=white, } ]
table[x index = 0, y index = 5] {data/ds_cuda_gtx770_p1u001.txt};
\addlegendentry{Kepler}

\addplot+[ black, only marks, mark=triangle*, mark options={ fill=white } ]
table {data/ds_cuda_gtx850M.txt};
\addlegendentry{Maxwell}

\addplot+[red, smooth, no marks, domain=0:31] {-16*x*x+1040*x+57024};

\end{axis}
\end{tikzpicture}

\caption{\label{fig:double-loop}Double loop timings corresponding to kernel in
Listing~\ref{lst:double-loop}. }

\end{figure}
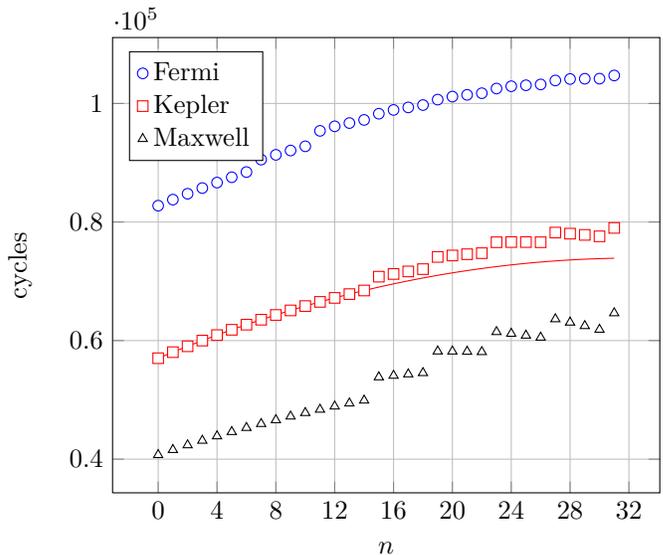

On our second test we used a kernel with nested loops, again the loops upper
bounds were set individually for each thread of the warp (see
Listing~\ref{lst:double-loop}). We used same patterns as for the single loop
(Table~\ref{tab:single-loop}) setting both loops bounds to the same value
($M=N$). The timings are presented in the Figure~\ref{fig:double-loop}. We can
observe a similar pattern as in the single loop case. The number of cycles at
first increases smoothly with the number of divergent threads, and then
exhibits "jumps" every four steps.

\section{CUDA divergence model} \label{sec:model}

To understand the observed behavior it is necessary to find out the details of
the CUDA thread divergence/re-convergence model. This is not explicitly stated
by NVIDIA apart from brief mention of the branch synchronization stack in
\cite{Tesla}. A more detailed description can be found in references
\cite{CudaHandbook, Barra, Branchement, gpgpusim} and \cite{OnCorrectness}.
There it was established that the CUDA implementation follows the approach
described in US patents \cite{Patent} and \cite{Patent2}. In here we briefly
describe this algorithm. Our description is based on reference \cite{Patent2}.
We choose reference \cite{Patent2} over \cite{Patent} because of the
\lstinline!SSY! instruction specification. While in \cite{Patent} it is
described as taking no arguments, \lstinline!cuobjdump! listing shows that it
expects one argument which matches the description in \cite{Patent2}. We
concentrate only on one type of control instruction which is relevant to our
example: the predicate branch instruction, ignoring all others.

Scalar Multiprocessor (SMX) processor maintains for each warp an \emph{active
mask} that indicates which threads in warp are active. When about to execute a
potentially diverging instruction (branch) compiler issues one
set-synchronization \lstinline!SSY! instruction. This instruction causes new
\emph{synchronization token} to be pushed on the top of synchronization stack.
Such token consist of three parts: mask, id and program counter (pc) and takes
64 bits - 32 of which are taken by the mask. In case of the \lstinline!SSY!
instruction the mask is set to active mask, id to \lstinline!SYNC! and pc is
set to the address of the synchronization point. The actual divergence is
caused by the \emph{predicated branch instruction} which has form
\lstinline!@P0 BRA! \emph{label} which translates to the pseudocode in
Algorithm~\ref{alg:predicate}. The \code{P0} is a 32bit \emph{predicate
register} that indicates which threads in the warp should execute (take) the
branch.

\begin{algorithm}

\begin{algorithmic}[1]
\If{None active threads take the branch}
  \State pc $\gets$ pc+1
\Else
  \If{ !(All active threads take the branch)}
    \State token.mask $\gets$ active\_mask \&\& !P0
    \State token.id $\gets$ DIV
    \State token.pc $\gets$ pc+1
    \State push(token)
  \EndIf
  \State active\_mask $\gets$ active\_mask \&\& P0
  \State pc $\gets$ label
\EndIf
\end{algorithmic}

\caption{\label{alg:predicate}Algorithm for executing a predicated branch
instruction \hbox{\lstinline!@P0 BRA! \emph{label}}.}

\end{algorithm}

The instructions are executed according to the pseudocode in the
Algorithm~\ref{alg:instruction}. The instruction may have a \emph{pop-bit set},
also denoted as \emph{synchronization command}, indicated by the suffix
\lstinline!.S! in the assembler output. The instruction which is suffixed with
\code{.S} will be called a \emph{carrier} instruction. When encountered it
signals the \emph{stack unwinding}. The token is popped from the stack and used
to set the active mask and the program counter. The reference \cite{Patent2}
does not specify exactly when carrier instructions are executed, before or
after popping the stack, so we have assumed that this is done after unwinding
the stack. Actually in both kernels that we have studied this carrier
instruction is \lstinline!NOP!, but this is not necessarily the case in general.

Summarizing: \lstinline!SSY! and predicated branch instructions (only if some
active threads diverge) push token onto the stack and the synchronization
command pops it.

\begin{algorithm}

\begin{algorithmic}[1]
\State Fetch the instruction
\If{Instruction is a  {\tt SSY} {\it label} instruction}
\State token.mask $\gets$ active\_mask
\State token.id $\gets$ SYNC
\State token.pc $\gets$ label
\State pc $\gets$ pc+1
\ElsIf{Instruction is a predicated branch instruction}
  \State Execute the instruction according to Algorithm~\ref{alg:predicate}
\Else
  \If{Is pop-bit set in instruction}
    \State token $\gets$ pop()
    \State active\_mask $\gets$ token.mask
    \State pc $\gets$ token.pc
    \State Execute the instruction
  \Else
    \State Execute the instruction
    \State pc $\gets$ pc+1
  \EndIf
\EndIf
\State goto 1
\end{algorithmic}

\caption{\label{alg:instruction}Algorithm for executing a CUDA instruction.}

\end{algorithm}


\section{Analysis}

We will try to understand the results of benchmarks from
Section~\ref{sec:timings} in view of the implementation described in
Section~\ref{sec:model}. We begin with disassembling the kernels. The results
are presented in Listings~\ref{fig:single-loop-disassembly} and
\ref{fig:double-loop-disassembly}. We have included only the relevant portions
of the assembler code.

Looking at the listing~\ref{fig:single-loop-disassembly}
\begin{lstlisting}[float,
  label=fig:single-loop-disassembly,
  caption={Disassembly of the single loop kernel}]
/*0060*/         ISETP.LT.AND P0, PT, R5, 0x1, PT;
                 ...
/*00e0*/         SSY 0x128;
/*00e8*/     @P0 BRA 0x120;
/*00f0*/         NOP;
/*00f8*/         NOP;
/*0100*/         IADD R4, R4, 0x1;
/*0108*/         FADD32I R0, R0, 1.3332999944686889648;
/*0110*/         ISETP.LT.AND P0, PT, R4, R5, PT;
/*0118*/     @P0 BRA 0x100;
/*0120*/         NOP.S;
/*0128*/         S2R R5, SR_CLOCKHI;
\end{lstlisting}
we notice all the instructions described in the previous section. To better
understand what is happening here we will make a "walk-through" this listing
assuming $n=2$.

\begin{figure}
  \centering\includestandalone{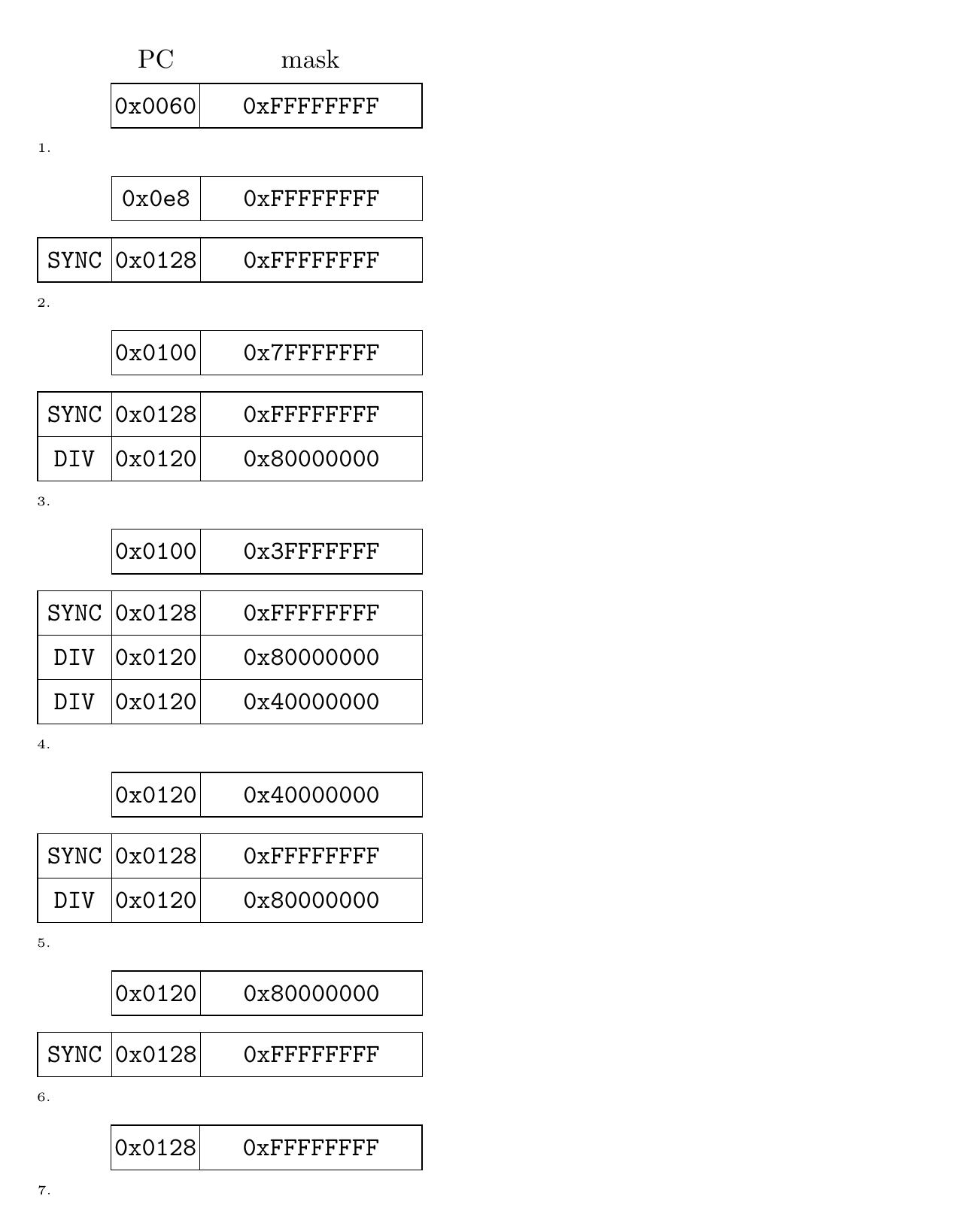}
\caption{\label{fig:walk-through}Walk through the single loop ($n=2$) showing
the stack history.} \end{figure}

\newcommand{\wfig}[1]{Figure~\ref{fig:walk-through}.#1}

We start on line 1, all bits in the active mask are set and the stack is empty
(\wfig1). The predicate register \code{P0} is set to the result of the
comparison $M<1$. In our test this condition is never fulfilled so all bits of
\code{P0} are cleared.

As line 4 contains a potentially diverging instruction a \lstinline!SSY!
instruction is issued on line 3 with address pointing past the loop to the line
12. The active mask and the address are pushed onto the stack (\wfig2).

On line 4 we have a predicated branch instruction, but as the register
\code{P0} is all false, this does not produce any divergence, and so does not
change the active mask and does not push any token on the stack. The control
then moves on to the next instructions.

Ignoring two \code{NOP}s the next instruction on line 7 increments the loop
counter $i$ in register \code{R4}, which was set to zero by an instruction not
shown on the listing. Next instruction on line 8 performs the addition.

On line 9 register \code{P0} is set to the result of testing the loop condition
$i<M$. On next line this register is used to predicate the branch instruction.
As $i=1$ all bits in \code{P0} are set and the branch is taken by all threads
and no token is pushed onto the stack. All the threads are active. The control
jumps to line 7 when the loop counter is again incremented. This continues
until $i=30$ without any changes to the stack.

When $i=30$ the condition on line 9 fails for the last thread of the warp and
last bit of the \code{P0} is cleared (\hbox{\code{P0=0x7FFFFFFF}}). The branch
on line 10 instruction is now diverging so a DIV token is pushed on the stack
with mask set to \code{\!P0=0x80000000} (not taken mask) and address equal to
PC+1 (line 11) (\wfig3). The active mask is set to \code{P0} which disables the
thread 31 and the execution jumps to line 7. Then the loop counter $i$ is
incremented to 31, but only in the active threads.

At line 9 the as $i=31$ the comparison fails for thread 30. The predicate
register \code{P0} will have its bit 30 cleared (\code{P0=0x3FFFFFFF}) so not
all active threads will take the branch on line 10. Again a DIV token will by
pushed on the stack with mask set to not taken mask (\code{0x7FFFFFFF &
\!0x3FFFFFFF = 0x40000000}) and \code{PC=0x0120}. Then the active mask is set
to \code{P0=0x3FFFFFFF} disabling threads 30 and 31 and control jumps to line 7
(\wfig4).

The loop counter is incremented again to 32 so all bits of the predicate
\code{P0} are cleared. There is no divergence in branch on line 10 as none of
the active threads take the branch and the control moves to line 11.

The \code{NOP} instruction on line 11 has pop-bit set, so a token is popped
from the stack. The active mask is set to the token mask and PC to the token's
PC. The NOP instruction is executed. And the control jumps again to line 11
that was the token program counter (\wfig5). Stack is popped again and after
executing the NOP instruction the control goes back to line 11 (\wfig6).

Then the last item is popped from the stack. This is the SYNC token that
restores the \code{0xFFFFFFFF} active mask and sets the PC to line 12, ending
the loop (\wfig7).

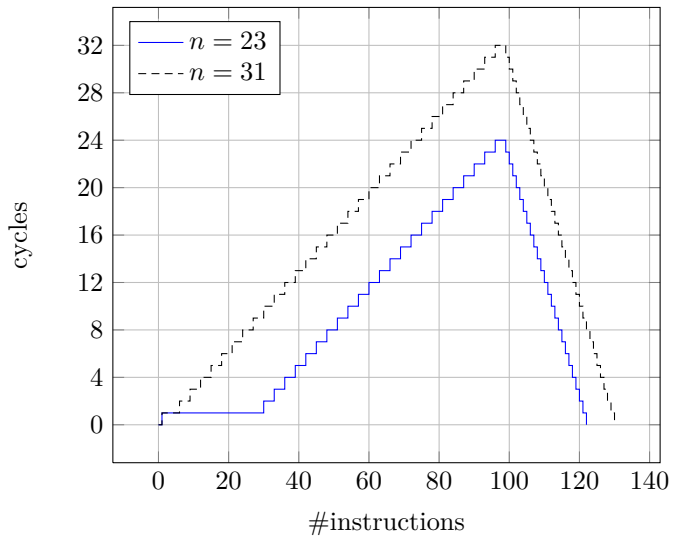
\begin{figure}
\centering
\begin{tikzpicture}
\begin{axis}[
  width=\columnwidth,
  xlabel={$\#$instructions},
  ylabel={cycles},
  ytick={0,4,8,12,16,20,24,28,32},
  grid=major,
  legend style= {
    legend pos=north west,
    legend cell align=left
  }
]

\addplot+[ const plot, no marks, blue, mark options={ fill=white } ]
table[x index = 1, y index = 4] {data/stack_ss_n23.txt};
\addlegendentry{$n=23$}

\addplot+[ const plot, no marks, black, mark options={ fill=white }, style =
densely dashed ]
table[x index = 1, y index = 4] {data/stack_ss_n31.txt};
\addlegendentry{$n=31$}

\end{axis}
\end{tikzpicture}

\caption{\label{fig:stack_ss}Re-convergence stack history for the single loop
kernel.}

\end{figure}

This walk-through lets us see the general pattern: as $n$ is increased we start
pushing \code{DIV} tokens on the stack earlier and earlier at each iteration
through the loop and the stack unwinding takes place at the end. The total
number of push/pop instructions is $n+1$ (the \code{SYNC} instruction always
pushes token on the stack) and the maximum stack size is also $n+1$. This is
illustrated on the Figure~\ref{fig:stack_ss} were we plot the stack history
showing the length of the stack as a function of the number of executed
instructions. This figure was obtained using a very simple emulator
implementing the algorithms 1 and 2.

We have also instrumented our code using CUPTI API from NVIDIA\cite{cupti}.
While this API cannot show the push/pop instructions, we have found out that
for $n=0\ldots15$ the number of executed instructions indeed increases by one
with increasing $n$ and that this is due to the increasing number of executions
of stack unwinding instruction \code{NOP.S} on line 11. This corresponds with
the linear raise of the number of cycles taken by the loop, as show in the
Figure~\ref{fig:single-loop}. By fitting to the first 16 points we have found
out that the penalty for one diverging branch is exactly 32 cycles on
\emph{Kepler} (see the continuous line in the Figure~\ref{fig:single-loop}) and
26 cycles on \emph{Maxwell}. Those 32 cycles include pushing the token on the
stack, executing a synchronization command and popping the stack. In
section~\ref{sec:detailed-timings} we will take a more detailed look at how
exactly those cycles are divided between those operations.

Looking again at the Figure~\ref{fig:single-loop} we notice that the first jump
occurs when length of the stack exceeds 16 \ie when $n>15$. We can assume that
this is the physical length of the fast on chip memory allocated for stack, and
growing the stack longer requires spilling the entries into some other memory
(as mentioned in \cite{Patent2}). After this jump the number of cycles again
increases by 32 at each step for the next four entries leading us to assume
that the entries are spilled four at the time e.g. in chunks of 32 bytes. This
is exactly the length of the L2 cache line suggesting that the entries are
spilled to the global memory. This would also explain why running a preheat
loop before measurements reduces the overall number of cycles taken by the
loops and makes the measurements deterministic, as it populates the cache.

We have assumed that during a push that would overflow the stack, the four
lowest entries are spilled into the global memory leaving the room for new four
highest entries. Likewise when popping the stack would require access to
elements that were spilled, they are loaded back from the memory. We have
implemented this scenario in our emulator. Comparing that with the timing
measurements we have found out that the spilling and loading back takes
together around 84 cycles on \emph{Kepler} architecture and 176 cycles on
\emph{Maxwell}.

Using the CUPTI API we have looked for additional load/store or cache hit/miss
instructions that would indicate the spill into the global memory, but found
none, however we have noticed appearance of some additional branch
instructions. Those instruction did not show up in the execution trace, so
could not be attributed to the particular instructions. The number of those
additional instructions was however exactly equal to the number of spills.
Without the detailed knowledge of NVIDIA micro-architecture it is impossible to
ascertain the meaning of those instructions. One plausible explanation would be
that the branch instructions are reissued after the stack content is spilled to
memory.

\begin{figure}
\centering
\begin{tikzpicture}
\begin{axis}[
  width=\columnwidth,
  xlabel={$\#$instructions},
  ylabel={cycles},
  ytick={0,4,8,12,16,20,24,28,32},
  grid=major,
  legend style= {
    legend pos=north west,
    legend cell align=left
  }
]

\addplot+[ const plot, no marks, blue, mark options={ fill=white } ]
table[x index = 1, y index = 4] {data/stack_ds_n23.txt};
\addlegendentry{$n=23$}

\addplot+[ const plot, no marks, black, mark options={ fill=white }, style =
densely dashed ]
table[x index = 1, y index = 4] {data/stack_ds_n31.txt};
\addlegendentry{$n=31$}

\end{axis}
\end{tikzpicture}

\caption{\label{fig:stack_ds}Re-convergence stack history for the double loops
kernel.}

\end{figure}
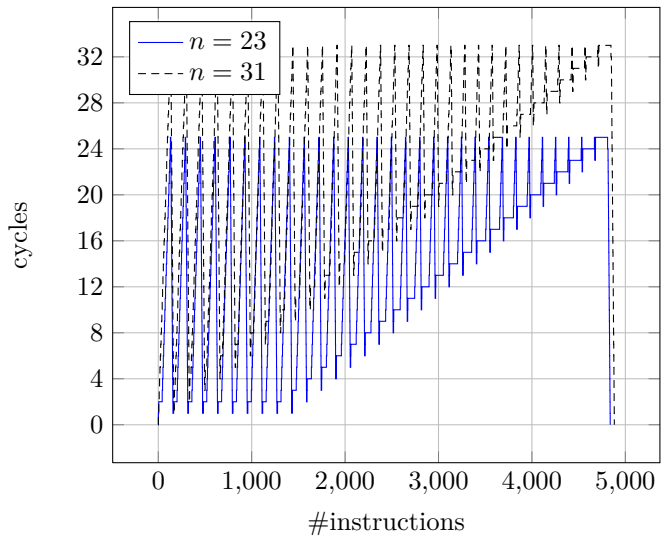

We did same analysis for the double loop kernel using the disassembly presented
in Listing~\ref{fig:double-loop-disassembly}.
\begin{lstlisting}[
  caption={Disassembly of the double loop kernel.},
  label={fig:double-loop-disassembly}]
/*0060*/         ISETP.LT.AND P0, PT, R8, 0x1, PT;
                 ...
/*0118*/         SSY 0x1a0;
/*0120*/     @P0 BRA 0x198;
/*0128*/         MOV R6, RZ;
/*0130*/         ISETP.LT.AND P0, PT, R9, 0x1, PT;
/*0138*/         MOV R7, RZ;
/*0140*/         SSY 0x178;
/*0148*/     @P0 BRA 0x170;
/*0150*/         IADD R7, R7, 0x1;
/*0158*/         FADD32I R0, R0, 1.3332999944686889648;
/*0160*/         ISETP.LT.AND P0, PT, R7, R9, PT;
/*0168*/     @P0 BRA 0x150;
/*0170*/         NOP.S;
/*0178*/         IADD R6, R6, 0x1;
/*0180*/         FADD32I R0, R0, 2.3333001136779785156;
/*0188*/         ISETP.LT.AND P0, PT, R6, R8, PT;
/*0190*/     @P0 BRA 0x130;
/*0198*/         NOP.S;
/*01a0*/         S2R R7, SR_CLOCKHI;
\end{lstlisting}
This is a  more complicated case as illustrated by the stack history
presented in the Figure~\ref{fig:stack_ds}.
The maximum length of the stack is now $n+2$ and we have found out that the
number of push operations is $x\cdot(65-x)/2 + 33$ (see the continuous line on
Figure~\ref{fig:double-loop}).

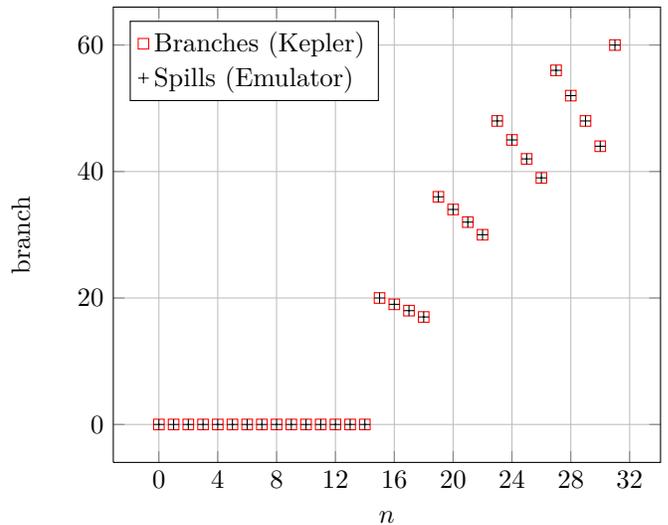
\begin{figure}
\centering
\begin{tikzpicture}
\begin{axis}[
  width=\columnwidth,
  xlabel={$n$},
  ylabel={branch},
  grid=major,
  xtick={0,4,8,12,16,20,24,28,32},
  legend style= {
    legend pos=north west,
    legend cell align=left
  }
]

\addplot+[ red, only marks, mark=square*, mark options={ fill=white, } ]
table[x index = 0, y expr = \thisrowno{10}-1089]
{data/ds_cuda_gtx770_p0u001.txt};
\addlegendentry{Branches (Kepler)}

\addplot+[ black, only marks, mark=+, mark options={ fill=white, } ]
table[x index = 0, y expr = \thisrowno{2}]{data/ds_cemu_no_first.txt};
\addlegendentry{Spills (Emulator)}

\end{axis}
\end{tikzpicture}

\caption{\label{fig:branch}Number of additional branch instructions issued in
the double loop kernel.}

\end{figure}

As in the single loop case we have found out that additional number of branch
instructions is executed and this number is exactly predicted by the number of
spills (see Figure~\ref{fig:branch}).


\begin{figure}
\centering
\begin{tikzpicture}
\begin{axis}[
  width=\columnwidth,
  xlabel={$n$},
  ylabel={cycles},
  grid=major,
  xtick={0,4,8,12,16,20,24,28,32},
  legend style= {
    legend pos=north west,
    legend cell align=left
  }
]

\addplot+[ red, only marks, mark=square*, mark options={ fill=white, } ]
table[x index = 0, y index = 1] {data/ds_diff.txt};

\end{axis}
\end{tikzpicture}

\caption{\label{fig:double-loop-diff}Difference between the actual and
predicted number of cycles for the double loop kernel on Kepler (GTX 770)
architecture.}

\end{figure}
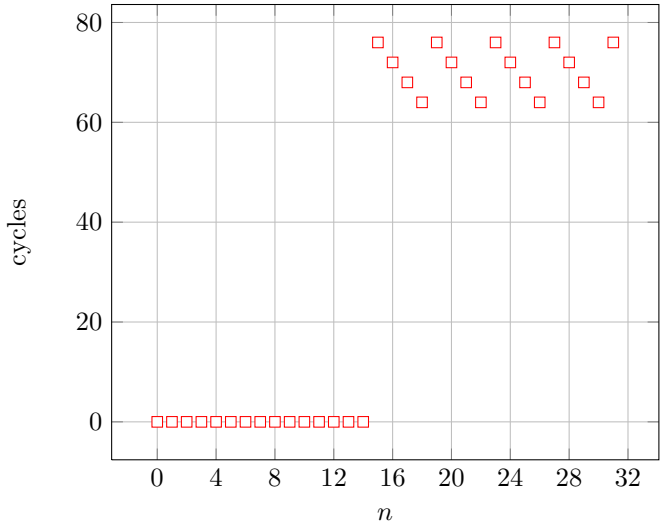

Using the parameters obtained from the single loop measurements we used our
emulator to predict the timings of the double loop kernel. The results are
presented in the Figure~\ref{fig:double-loop-comparison}. The agreement while
not perfect is still very good (76 cycles difference in the worst case which
amounts to $~0.3\%$). Looking at the Figure~\ref{fig:double-loop-diff} we see
that the difference between the actual and predicted number of cycles follows a
very regular pattern and happens only on spill occurrence, suggesting some
simple mechanism contributing to the exact number of cycles needed to perform
stack spill depending on previous spills history.


\begin{figure}
\centering
\begin{tikzpicture}
\begin{axis}[
  width=\columnwidth,
  xlabel={$n$},
  ylabel={cycles},
  grid=major,
  xtick={0,4,8,12,16,20,24,28,32},
  legend style= {
    legend pos=north west,
    legend cell align=left
  }
]

\addplot+[ red, only marks, mark=square*, mark options={ fill=white, } ]
table[x index = 0, y index = 5] {data/ds_cuda_gtx770_p1u001.txt};
\addlegendentry{Kepler}

\addplot+[ black, only marks, mark=triangle*, mark options={ fill=white, } ]
table[x index = 0, y index = 1] {data/ds_cuda_gtx850M.txt};
\addlegendentry{Maxwell}

\addplot+[ black, only marks, mark=+, mark options={ fill=white, } ]
table[x index = 0, y expr = \thisrowno{1}-1056+57024]
{data/ds_cemu_no_first.txt};
\addlegendentry{Emulator}
\addplot+[ black, only marks, mark=+, mark options={ fill=white, } ]
table[x index = 0, y expr = \thisrowno{1}-858+40716.91]
{data/ds_cemu_maxwell.txt};

\addplot+[red, smooth, no marks, domain=0:31] {-16*x*x+1040*x+57024};

\end{axis}
\end{tikzpicture}

\caption{\label{fig:double-loop-comparison}Comparison of the double loop
timings with emulator results. }

\end{figure}
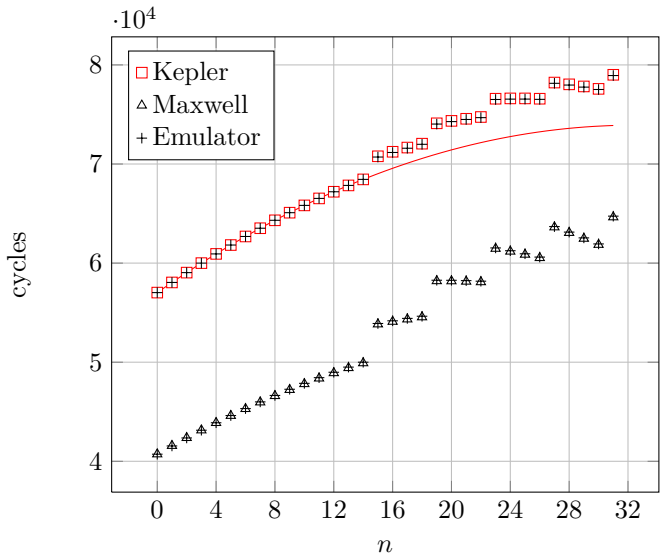


\subsection{Detailed timings}
\label{sec:detailed-timings}

In the previous section we have calculated the costs associated with diverging
predicated branch instructions. That was the total cost associated with pushing
and popping the stack as well as executing the synchronization command carrier
instruction. In this section we are attempting a more detailed view. To this
end we will use the kernel from listing~\ref{lst:single-loop-precise}.
\begin{lstlisting}[linerange={8-31}, label={lst:single-loop-precise},  float,
caption={Single loop kernel with detailed timings.}]
__global__ void single_loop_precise(int *limits, float *out, llong *timer) {

  int tid = blockDim.x * blockIdx.x + threadIdx.x;
  int M = limits[threadIdx.x];
  float sum = out[tid];

#pragma unroll
  for (int k = 0; k < N_PREHEAT; k++)
    for (int i = 0; i < M; i++) {
      sum +=1.3333f;
    }

  timer[tid] = clock64();

  for (int i = 0; i < M; i++) {
    sum += 1.3333f;
    timer[(i + 1) * 32 + tid] = clock64();
  }

  __syncthreads();
  timer[33 * warpSize + tid] = clock64();

  out[tid] = sum;
}
\end{lstlisting}
This is essentially the single loop kernel with additional timing instruction
inserted \emph{inside} the loop (line 17). According to the walk-through in the
previous section those instructions should be executed \emph{before} unwinding
the stack, while the last instruction on line 21 should be executed after.

At first we did not include the synchronization command on line 20, as we
deemed it unnecessary within one warp. The results were however not compatible
with the previous ones. Inspection of the disassembly in
listing~\ref{fig:precise-no-sync} reveled that actually all the timing
instructions were executed before the stack unwinding triggered by the
synchronization command on line 34.

\begin{lstlisting}[float,
  caption={Disassembly of the  single loop kernel with detailed
           timings without synchronization.},
  label={fig:precise-no-sync}]
/*0068*/         S2R R9, SR_CLOCKHI;
/*0070*/         IADD R4.CC, R8, R8;
/*0078*/         ICMP.LT R5, R5, R9, R8;
/*0080*/         IADD.X R5, R5, R5;
/*0088*/         MOV32I R9, 0x8;
/*0090*/         IMAD R8.CC, R0, R9, c[0x0][0x30];
/*0098*/         IMAD.HI.X R9, R0, R9, c[0x0][0x34];
/*00a0*/         ST.E.64 [R8], R4;
/*00a8*/         S2R R5, SR_CLOCKHI;
/*00b0*/         S2R R8, SR_CLOCKLO;
/*00b8*/         S2R R9, SR_CLOCKHI;
/*00c0*/         IADD R4.CC, R8, R8;
/*00c8*/         ICMP.LT R5, R5, R9, R8;
/*00d0*/         IADD.X R5, R5, R5;
/*00d8*/         ISETP.LT.AND P0, PT, R7, 0x1, PT;
/*00e0*/         MOV R10, RZ;
/*00e8*/         SSY 0x178;
/*00f0*/     @P0 BRA 0x170;
/*00f8*/         MOV32I R11, 0x8;
/*0100*/         IADD R10, R10, 0x1;
/*0108*/         FADD32I R6, R6, 1.333;
/*0110*/         ISCADD R9, R10, R0, 0x5;
/*0118*/         IMAD R8.CC, R9, R11, c[0x0][0x30];
/*0120*/         IMAD.HI.X R9, R9, R11, c[0x0][0x34];
/*0128*/         ST.E.64 [R8], R4;
/*0130*/         S2R R5, SR_CLOCKHI;
/*0138*/         S2R R8, SR_CLOCKLO;
/*0140*/         S2R R9, SR_CLOCKHI;
/*0148*/         IADD R4.CC, R8, R8;
/*0150*/         ICMP.LT R9, R5, R9, R8;
/*0158*/         IADD.X R5, R9, R9;
/*0160*/         ISETP.LT.AND P0, PT, R10, R7, PT;
/*0168*/     @P0 BRA 0x100;
/*0170*/         IADD.S R0, R0, 0x420;
\end{lstlisting}

Adding the \code{__syncthreads()} instruction on line 20 solved the problem.
The overall results now matched the results from the previous section. Of
course the total times differed substantially as the code had now extra
instructions, but the differences in number of cycles between the runs with
various values of $n$ were identical for $n\le15$ (no spills) and differed by
few cycles for $n>15$.

We have gathered the timelines for each $n$ and each thread in warp but we used
only the timings for thread 0 -- the one that always iterates 32 times through
the loop. We have found out that when there were no spills ($n\le15$) then the
total cost was last code segment: the unwinding of the stack. There was no cost
associated with predicated brach instruction \ie the push operation. This may
be explained by the fact that the latency of push operation can be hidden as
the token pushed onto the stack is not needed by the next instruction. While
unwinding the stack however the popped token is used immediately to set active
mask and PC. Additionally the carrier instruction has to be executed, even if
this is a \code{NOP} it still has to be fetched an decoded.

When $n$ was greater then $15$, \ie the stack was spilled in the memory, the
cost were divided between the branch instruction that now took around 40 cycles
longer and stack unwinding that now took around 42 cycles longer. This
corresponds well with the overall 84 cycles penalty estimated in the previous
section.


\section{Summary}

The Single Thread Multiple Threads (SIMT) programming model introduced by
NVIDIA with CUDA technology gives the programmer an illusion of writing a
multithreaded application. In reality the threads are executed in groups of 32
threads (\emph{warps}) performing essentially vector operations. The illusion
of multithreading is maintained by an elaborate mechanism for managing the
divergent threads within a warp. This mechanism is completely hidden from the
programmer and no description is provided by NVIDIA in their programming guide.
In this contribution we have presented a detailed study of two CUDA kernels
leading to a large thread divergence. Following other sources, we have assumed
that the management of the diverging threads is based on a synchronization
stack and follows closely the idea described in reference \cite{Patent2} which
is a NVIDIA owned patent. That assumption allowed us to fit the observed
execution times with accuracy better then $1\%$.

We have estimated the cost of a diverging branch instruction to be
\emph{exactly} 32 cycles on \emph{Kepler} architecture provided that the
maximum stack length did not exceed 16. After that stack entries had to be
spilled into memory which carried an additional penalty of $\sim84$ cycles. On
\emph{Maxwell} the estimated cost of branch divergence was considerably
shorter: only $24$ cycles, but the cost of spilling was much higher $\sim176$
cycles. More detailed timings are suggesting that all of that cost, in case of
no spills, can be attributed to stack unwinding.

We also performed the measurements on \emph{Fermi} architecture, there the
pattern seems to be slightly different suggesting that the physical thread
synchronization stack length is shorter in this architecture.

We have analyzed only a very specific case: loops with different bounds across
the threads, this is however a quite natural scenario. We would like to
emphasize that in this case the additional costs reported by us are the costs
of the divergence management mechanism alone, excluding additional penalty of
incomplete vectorization. Nevertheless they implied an overhead of $\sim 20\%$
in case of 14 diverging threads for the double loop kernel and $\sim 25\%$ for
the single loop. In the more general case there would be additional costs
associated with the serialization of the different branches.

Our work was confined to NVIDIA CUDA technology it would be however interesting
how this corresponds to shader programming both in DirectX and OpenGL and we
are planning to extend our work in this direction in the future.


\bibliographystyle{IEEEtran}
\bibliography{IEEEabrv,divergence}

\begin{thebibliography}{10}
\providecommand{\url}[1]{#1}
\csname url@samestyle\endcsname
\providecommand{\newblock}{\relax}
\providecommand{\bibinfo}[2]{#2}
\providecommand{\BIBentrySTDinterwordspacing}{\spaceskip=0pt\relax}
\providecommand{\BIBentryALTinterwordstretchfactor}{4}
\providecommand{\BIBentryALTinterwordspacing}{\spaceskip=\fontdimen2\font plus
\BIBentryALTinterwordstretchfactor\fontdimen3\font minus
  \fontdimen4\font\relax}
\providecommand{\BIBforeignlanguage}[2]{{%
\expandafter\ifx\csname l@#1\endcsname\relax
\typeout{** WARNING: IEEEtran.bst: No hyphenation pattern has been}%
\typeout{** loaded for the language `#1'. Using the pattern for}%
\typeout{** the default language instead.}%
\else
\language=\csname l@#1\endcsname
\fi
#2}}
\providecommand{\BIBdecl}{\relax}
\BIBdecl

\bibitem{Reconstruction}
P.~Bia{\l}as, J.~Kowal, A.~Strzelecki \emph{et~al.}, ``{GPU} accelerated image
  reconstruction in a two-strip {J-PET} tomograph,'' \emph{arXiv preprint
  arXiv:1502.07478}, 2015.

\bibitem{Tesla}
E.~Lindholm, J.~Nickolls, S.~Oberman, and J.~Montrym, ``{NVIDIA} {Tesla}: A
  unified graphics and computing architecture,'' \emph{IEEE Micro}, vol.~28,
  no.~2, pp. 39--55, 2008.

\bibitem{Demystifying}
H.~Wong, M.-M. Papadopoulou, M.~Sadooghi-Alvandi, and A.~Moshovos,
  ``Demystifying {GPU} microarchitecture through microbenchmarking,'' in
  \emph{Performance Analysis of Systems \& Software (ISPASS), 2010 IEEE
  International Symposium on}.\hskip 1em plus 0.5em minus 0.4em\relax IEEE,
  2010, pp. 235--246.

\bibitem{CudaHandbook}
N.~Wilt, \emph{The {CUDA} Handbook: A Comprehensive Guide to {GPU}
  Programming}.\hskip 1em plus 0.5em minus 0.4em\relax Pearson Education, 2013.

\bibitem{Barra}
S.~Collange, M.~Daumas, D.~Defour, and D.~Parello, ``{Barra}: A parallel
  functional simulator for {GPGPU},'' in \emph{Modeling, Analysis \& Simulation
  of Computer and Telecommunication Systems (MASCOTS), 2010 IEEE International
  Symposium on}.\hskip 1em plus 0.5em minus 0.4em\relax IEEE, 2010, pp.
  351--360.

\bibitem{Branchement}
------, ``Comparaison d'algorithmes de branchements pour le simulateur de
  processeur graphique {Barra},'' in \emph{13`eme Symposium sur les
  Architectures Nouvelles de Machines}, 2009, pp. 1--12.

\bibitem{gpgpusim}
\BIBentryALTinterwordspacing
T.~M. Aamodt \emph{et~al.}, ``{GPGPU-Sim},'' 2012. [Online]. Available:
  \url{http://www.gpgpu-sim.org}
\BIBentrySTDinterwordspacing

\bibitem{OnCorrectness}
A.~Habermaier and A.~Knapp, ``On the correctness of the {SIMT} execution model
  of {GPU}s,'' in \emph{Programming Languages and Systems}.\hskip 1em plus
  0.5em minus 0.4em\relax Springer, 2012, pp. 316--335.

\bibitem{Patent}
\BIBentryALTinterwordspacing
B.~W. Coon and J.~E. Lindholm, ``System and method for managing divergent
  threads in a {SIMD} architecture,'' US Patent US7\,353\,369 B1, 04 01, 2008.
  [Online]. Available: \url{https://www.google.com/patents/US7353369}
\BIBentrySTDinterwordspacing

\bibitem{Patent2}
\BIBentryALTinterwordspacing
B.~W. Coon, J.~R. Nickolls, L.~Nyland, P.~C. Mills, and J.~E. Lindholm,
  ``Indirect function call instructions in a synchronous parallel thread
  processor,'' US Patent US8\,312\,254 B2, 11 13, 2012. [Online]. Available:
  \url{https://www.google.com/patents/US8312254}
\BIBentrySTDinterwordspacing

\bibitem{cupti}
\BIBentryALTinterwordspacing
NVIDIA, ``{CUPTI}: {CUDA} toolkit documentation,'' 2014. [Online]. Available:
  \url{http://docs.nvidia.com/cuda/cupti/}
\BIBentrySTDinterwordspacing

\end{thebibliography}

\end{document}